\documentclass[12pt,aps]{revtex4}
\usepackage{epsfig,amsmath,amssymb}

\newcommand{\bea}{\begin{eqnarray}}
\newcommand{\eea}{\end{eqnarray}}

\newcommand{\dbar}{d\hskip -0.4em ^-}

\begin{document}

\title{Electric dipole moment of the neutron in \\
  Two Higgs Doublet Models with flavor changing}

\vspace{0.5cm}

\author{Jan O. Eeg}

\email{j.o.eeg@fys.uio.no}
\affiliation{Department of Physics, University of Oslo,
P.O.Box 1048 Blindern, N-0316 Oslo, Norway}

\vspace{0.5 cm}

\begin{abstract}

I consider  contributions to the neutron electric dipole moment within 
 Two Higgs Doublet Models which allow for small flavor changing 
neutral Higgs couplings.
In a previous paper, I considered  flavor changing interactions for  
the Standard Model  Higgs boson to first order in the flavor changing
 coupling. In that paper 
I found that the obtained value of 
the neutron electric dipole moment 
 were below the present
 experimental limit, given previous restrictions on such couplings.
Because this was an effective theory, the result depended on
 an ultraviolet cut off $\Lambda$,
 parametrized as $ln(\Lambda^2)$.

In the present paper I demonstrate that, when going to Two Higgs Doublet Models,
 the result stays  the same as in the previous paper, 
up to  $M_{SM}^2/M_H^2$ corrections, where 
$M_{SM}$ is the  mass of the  top-quark
 or the  $W$-boson. $M_H$ is the mass of the heavy neutral 
scalar Higgs-boson $H$ which is much heavier than the  light
 neutral  Higgs boson $h$ with mass $M_h$. 
In the limit $M_H^2 \gg M_h^2$,
  the $ln(\Lambda^2)$
 behaviour in the previous paper is replaced
 by $ln(\widetilde{M_H}^2)$, where $\widetilde{M_H}$ is of order $M_H$.

I also explain how some divergences due to exchange of the pseudoscalar 
Higgs $A$ are cancelled by similar contributions from the scalar 
heavy Higgs $H$, and that  these contributions, and finite contributions from
$A$-exchange,  are suppressed.

\end{abstract}

\maketitle

\vspace{1cm}


\section{Introduction}

 Studies of CP-violationg phenomena  are important in order to understand 
the asymmetry between matter and anti-matter
in the universe. Electric dipole moments (EDMs) of elementary particles
violates time reversal symmetry. Assuming CPT symmetry to be valid,
electric dipole moments (EDMs) are therefore also CP-violating.  
EDMs of elementary particles has not been seen yet.
 But experiental searches for EDMs and theoretical studies of EDMs are important
 because  any measured result  bigger than the  tiny values obtained within
Standard Model (SM) 
 will signal New Physics. Reviews on EDMs within the SM and beyond are given
 in \cite{Pospelov:2005pr,Fukuyama:2012np,Dekens:2014jka,Jung:2013hka}.
 For the EDM of
 the neutron (nEDM = $d_n$) discussed in this paper, the present
 experimental bound is
\cite{Baker:2006ts} 
\begin{equation}
d_n^{exp}/e   \le 2.9 \times 10 ^{-26} \,  \mbox{cm} \; .
\label{dn-bound}
\end{equation}
The corresponding values calculated within the SM  are   ranging from
 $10^{-34} e $ cm to $10 ^{-31} \, e$~cm, depending on the
 considered mechanism \cite{Shabalin:1980tf,Czarnecki:1997bu,Nanopoulos:1979my,Morel:1979ep,Gavela:1981sk,Eeg:1982qm,Eeg:1983mt,Khriplovich:1981ca,McKellar:1987tf,Mannel:2012qk}. Many models beyond the SM (BSM) gives bigger values compared to those obtained within the SM
\cite{Pospelov:2005pr,Fukuyama:2012np,Dekens:2014jka,Jung:2013hka,Maiezza:2014ala,Buchmuller:1982ye,Manohar:2006ga,Altmannshofer:2010ad,Buras:2010zm,Degrassi:2010ne,Arnold:2012sd,Brod:2013cka,He:2014uya,Bertolini:2019out,Fuyuto:2015ida}.
 In the presence  of New Physics, flavor 
physics may also give
 useful CP-violating observables. These may occur for instance in
CP-violating  mesonic decays
\cite{Altmannshofer:2010ad,Buras:2010zm,Altmannshofer:2012ur}.
 Within New Physics models describing  such
 processes there might be
mechanisms that also generate new contributions to the electric dipole moments
 of quarks (see e.g. \cite{Fajer:2014ara}). 

The electric dipole moment of a single fermion  has the form
\begin{equation}
{\cal L}_\mathrm{fEDM} \; = \; \frac{i}{2} \, d_f \,
\bar{\psi_f} \sigma_{\mu \nu} \, F^{\mu \nu} \, \, \gamma_5 \psi_f
\; \; ,
\label{effLag}
\end{equation}
where $d_f$ is the electric dipole moment of the fermion,
$\psi_f$ is the fermion (quark) field, $F^{\mu \nu}$ is the electromagnetic 
field tensor,  and
$\sigma_{\mu \nu} = i [\gamma_\mu, \, \gamma_\nu]/2$ is the dipole operator
 in Dirac space.

We still do not know in detail the   properties and couplings
of the Higgs boson. For instance, the SM
 Higgs might mix with (a) higher mass scalar(s) in the BSM.
 It has been suggested \cite{Goudelis:2011un,Blankenburg:2012ex,Harnik:2012pb,Greljo:2014dka,Gorbahn:2014sha,Dorsner:2015mja}  that the
SM Higgs boson might have  small flavor changing (FC) couplings to 
fermions.  Such couplings may also
be CP-violating.  
Studying flavor changing processes 
like $K-\bar{K}, \,D-\bar{D} \,$, and $ B-\bar{B}$ - mixings, and 
also  leptonic flavor 
changing decays like $\mu \rightarrow e \, \gamma$
 and $\tau \rightarrow \mu \, \gamma$,  
 bounds on {\it quadratic} expressions of 
such flavor changing couplings was obtained. 
 For the leptonic cases  
 two loop diagrams of Barr-Zee type \cite{Barr:1990vd} for EDMs,
 were also considered
\cite{Goudelis:2011un,Blankenburg:2012ex,Harnik:2012pb,Chang:1993kw,Leigh:1990kf}. Extra couplings of the SM Higgs to quarks has also been considered 
in \cite{Brod:2018lbf}.

In a previous paper \cite{Eeg:2016fsy}, I extended the analysis of EDMs of 
light quarks  with
flavor changing SM Higgs couplings (FCH) to {\it two loop diagrams}.
Going from one loop to two loop diagrams there is a priori a loop suppression.
However, in general, it is known that some two loop diagrams might give
 bigger amplitudes than one loop diagrams
because of  helicity flip(s) in the latter
\cite{Blankenburg:2012ex,Harnik:2012pb,Chang:1993kw,Bjorken:1977vt}.
And further, what is most important in the present case,
 I calculated {\it two loop diagrams}
 containing a small
flavor changing SM Higgs coupling (FCH) {\em to first order only}, 
-in contrast to {\it one loop } contributions 
with a small FCH coupling to {\it second order}, as
 in \cite{Blankenburg:2012ex,Harnik:2012pb}.
Therefore, if the FCH couplings exist, my two loop amplitudes might
numerically  compete or even dominate over the one loop amplitudes calculated
previously.

Some of the two loop diagrams were divergent and parametrised by a ultraviolet
 cut-off $\Lambda$ \cite{Eeg:2016fsy}.
In the present paper I address the same diagrams within 
Two Higgs Doublet Models (2HDMs). In such renormalisable  models one knows that
the final result does not depend on divergent contributions. In section V I
 demonstrate how divergent terms disappear due to cancellations of
 different terms  in the general 2HDMs. And I find how the phenomenological
 FCH coupling of refs .\cite{Blankenburg:2012ex,Harnik:2012pb} is expressed
 within 2HDMs.

 For descriptions of 2HDMs, see the review by
 Branco et al.
\cite{Branco:2011iw}, and also more recent papers \cite{Altmannshofer:2016zrn,Botella:2016krk,Grzadkowski:2016szj,Alves:2018kjr,Grzadkowski:2018ohf,Ouazghour:2018mld,Hou:2019grj,Herrero-Garcia:2019mcy,deMedeirosVarzielas:2019dyu,Correia:2019vbn,Fuyuto:2019svr,Chen:2019pnt} .  Phenomenological consequences of 2HDMs
 are given in \cite{Crivellin:2013wna}.

Some technical details from the two loop calculations are given in the Appendix.

\section{Flavor Changing Physical Higgs?}

Within the framework in  \cite{Goudelis:2011un,Blankenburg:2012ex,Harnik:2012pb,Greljo:2014dka,Gorbahn:2014sha}
 the effective interaction Lagrangian for a flavor transition between 
fermions of the same charge 
 due to SM Higgs boson exchange might be obtained from a six dimensial
 non-renormalizable  
Higgs type Yukawa-like interactions as 
shown explicitly in \cite{Harnik:2012pb,Dorsner:2015mja} :
\begin{equation} 
  {\cal L}^{(D)}  \, =  \, - \, \lambda_{ij} \, (\overline{Q_L})_i \, \phi  
\, (d_R)_j \, - \, \frac{\tilde{\lambda}_{ij}}{\Lambda_{NP}^2} 
 \, (\overline{Q_L})_i \, \phi \, (d_R)_j \, (\phi)^\dagger \phi \, + \; h.c. \; ,
\label{FCNClad}
\end{equation}
where the  generation indices $i$ and $j$ running from 1 to 3 
are understood to be summed over; $i.e. \, $ $d_j = d,s,b$ for $j=1,2,3$.
Further,  $\phi$ is the SM Higgs $SU(2)_L$ doublet field, $(Q_L)_i$ are the
 left-handed $SU(2)_L$ 
quark doublets, and the  $(d_R)_j$'s are the right-handed $SU(2)_L$ singlet 
$d$-type quarks in a general basis. 
 Further, 
$\Lambda_{NP}$ is the scale where New Physics is assumed to appear.
There is a similar term  ${\cal L}^{(U)} $ like the one in (\ref{FCNClad}) for 
  right-handed type $u$-quarks, $u_j$, $i.e.$ $u_j =u,c,t$ for $j=1,2,3$.
If higher states from a 
renormalized theory are integrated out.
 interactions may occur like in (\ref{FCNClad}) below.

In such cases  the Yukawa interaction for the SM neutral Higgs boson $h^0$
 to $d$-type quarks has  the form
\begin{equation}
{\cal L}_Y^{(D)} \, = \, - \,  h^0 \, (\bar{d}_L)_i  
 \left( Y^{(D)}_R \right)_{ij} \, (d_R)_j \, + h.c. \;  ,
\label{FCH-coupl}
\end{equation}
where
\begin{equation}
  Y^{(D)}_R \, = \, 
\frac{{\cal M}^{(D)}}{v} \, - \, \epsilon^{(D)}_R 
\label{Y-coupl}
\end{equation}
Here ${\cal M}^{(D)}$ is the mass matrix for $d$-type quarks giving 
the SM coupling, and  $\epsilon^{(D)}_R$ is 
the  part which goes beyond the SM, related to the six dimensional operators
in (\ref{FCH-coupl}).
Explicitly, one finds \cite{Harnik:2012pb,Dorsner:2015mja}:
\begin{equation} 
({\cal M}^{(D)})_{ij} \, = \, \frac{v}{\sqrt{2}} \left(\lambda_{ij} \, + \,
 \frac{v^2 \tilde{\lambda}_{ij}}{2 \Lambda_{NP}^2}\right) \; \; ,
\; \; \mbox{and}  \; \; \,
(\epsilon^{(D)}_R)_{ij} \, = \, 
\frac{v^2 \tilde{\lambda}_{ij}}{\sqrt{2} \Lambda_{NP}^2} \; .
\label{FC-matr}
\end{equation} 
As usual $v$ is the vacuum value 246 GeV for the SM Higgs field.
The mass matrix ${\cal M}^{(D)}$ may be rotated  to diagonal form.
 However, this rotation will in general not give a diagonal $\epsilon^{(D)}_R$,
such that the SM Higgs coupling to fermions will in general  be flavor changing.
Thus, for $i \neq j$, $ Y^{(D)}_R = - \epsilon^{(D)}_R $.
In \cite{Goudelis:2011un,Blankenburg:2012ex,Harnik:2012pb,Greljo:2014dka,Gorbahn:2014sha,Dorsner:2015mja} bounds of FCH couplings to second order are obtaind
 from various flavor changing procecesses. In my own case, I will need the
 bound on $Y^{(D)}_R(d \rightarrow b)$
 from $B_d - \overline{B_d}$-mixing \cite{Harnik:2012pb}.

\section{Yukawa interactions for 2HDMs}

For  2HDMs the extended
 Yukawa interactions for right-handed type $d$-quarks may then, in the most 
general case,  be written as \cite{Alves:2018kjr} :
\begin{equation}
\, - \, {\cal L}_\Gamma^{(D)}  \, =   \, 
(\overline{(Q_L)_i}^0)^r \left[ (\Gamma_1)_{ij}^{rs} \, (\Phi_1)^s \, + \,  
(\Gamma_2)_{ij}^{rs} \,(\Phi_2)^s \right] \, (d_R)_j^0    + \; h.c. \; ,
\label{YukawaGendD}
\end{equation}
where $i,j$ are as before generation indices running from 1 to 3 and $r,s$ 
are $SU(2)_L$ indices running from 1 to 2. 
The upper index $0$
 denotes the fields before  diagonalization 
of the mass matrices in the quark sector.
Thus the $\Gamma$'s are $2 \times 2$ dimensional in $SU(2)_L$ space and 
$3 \times 3$ dimensional in generation space.
The fields $\Phi_{1,2}$ are the  two Higgs fields.

 For the  
right-handed type $u$-quarks one has similarly as (\ref{YukawaGendD}):
\begin{equation}
 \, - \,  {\cal L}_\Delta^{(U)}  \, =  \, \overline{(Q_L)}^0 \,\left[\Delta_1 \,
\tilde{\Phi}_1 \, + \,  \Delta_2 \,
\tilde{\Phi}_2 \right] \, (u_R)^0 \, + \; h.c. \; ,
\label{YukawaGenu}
\end{equation}
where the generation and $SU(2)_L$ indices are suppressed.
$\Gamma_{1,2}$ and $\Delta_{1,2}$ are in general complex and independent 
 quantities.
 In many papers one discusses restrictions 
on 2HDMs  to avoid flavor changing neutral currents completely.
 But in this paper the point is to study such potential effects.

The two Higgs doublets may for $n=1,2$ be written
 \cite{Branco:2011iw,Grzadkowski:2016szj} :
\begin{equation}
\Phi_n = \, e^{i \,  \xi_n} \,
\left( \begin{array}{cc} \phi_n^+ \\ \frac{1}{\sqrt{2}}(v_n + \rho_n + i \eta_n) 
 \end{array} \right)  \; ,  \quad  \tilde{\Phi}_n \;  = \, e^{-i \,  \xi_n} \, 
  \left( \begin{array}{cc} \frac{1}{\sqrt{2}}(v_n + \rho_n -  i \eta_n) 
\\ - (\phi_n^+)^\dagger \end{array} \right)  \; ,
\label{Higgses}
\end{equation}
where $\tilde{\Phi}_n \, \equiv i \sigma_2 \Phi_n^*$, and
 $ e^{i \,  \xi_n}$ are phase factors. One introduces 
the parameter $\beta$ through 
\begin{equation}
tan \beta \; \equiv \frac{v_2}{v_1} \; \, .
\label{beta}
\end{equation}
After diagonalisation of the mass matrix for the neutral fields $\rho_{1,2}$ 
obtained from the Higgs potential \cite{Branco:2011iw} one finds 
the  neutral scalar mass eigenstates
\begin{equation}
h \; = \; \rho_1 s_\alpha \, - \, \rho_2 c_\alpha \, \; ;  \quad
 H \; = \; -\rho_1 c_\alpha \, - \, \rho_2 s_\alpha \;  \, ,
\label{alpha1}
\end{equation}
and the  inverted relations are:
\begin{equation}
- \rho_1 \; = \;  
 - \, H \, c_\alpha \,-  s_\alpha \, h  \; \,  ; \quad  
 - \rho_2 \; = \; h \, c_\alpha \, - \, H\, s_\alpha \; \, .
\label{alpha2}
\end{equation}
Here $s_\alpha \equiv sin \alpha$ and $c_\alpha \equiv cos \alpha$,   
where $\alpha$ is the mixing angle  coming from the  diagonalisation
of the mass matrix
of the $\rho_{1,2}$ fields.
Note that in the previous paper \cite{Eeg:2016fsy} the SM Higgs 
was denoted $H$.
In the present  paper this symbol is reserved for the heavy neutral
 Higgs boson within 2HDMs.

In 2HDMs one often uses  the Higgs basis, where  the doublet 
fields $H_{1,2}$ are 
defined by
 \begin{equation}
e^{-i \, \xi_1} \, \Phi_1 \, = \, c_\beta \, H_1 \, + \, s_\beta \, H_2 \; ; \quad 
e^{-i \, \xi_2} \, \Phi_2 \, = \, s_\beta \, H_1 \, -  \, c_\beta \, H_2 \; , \; 
\label{Higgsbase}
\end{equation}
where $c_\beta \equiv cos \beta$ and $s_\beta \equiv sin \beta$. With this
 definition 
 $H_1$ has a vacuum value $v = \sqrt{v_1^2 + v_2^2}$
 and $H_2$ has zero vacuum value. Thus,  in  this basis
\begin{equation}
H_1  = \, 
\left( \begin{array}{cc} G^+ \\ \frac{1}{\sqrt{2}}(v + h^0 + i G^0) 
 \end{array} \right)  \; ,  \quad  H_2 \; = \, 
  \left( \begin{array}{cc} H^+ 
\\ \frac{1}{\sqrt{2}}(R^0 + i A) \end{array} \right)  \; ,
\label{SMHiggs}
\end{equation}
where  $v$ is the vacuum value 246 GeV for the SM Higgs field, and 
$G^+$ and $G^0$ are Goldstone fields.
$H^+$ is the charged Higgs field and $A$
the neutral pseudoscalar field within 2HDMs.
Now  the neutral scalar fields $h^0$ and $R^0$ can be written in terms of 
the physical (within 2HDMs) neutral scalars $h$ and $H$ as
 \begin{equation}
h^0  \, = \, - c_\theta \, H \, + \, s_\theta \, h \; \; ;  \; \; 
R^0\, = \, s_\theta \, H \, + \, c_\theta \, h \; , \; 
\label{HiggMix}
\end{equation}
where
\begin{equation}
 c_\theta \equiv cos \theta \; \, , \quad \mbox{and} \; \; 
 s_\theta \equiv sin \theta \; ,
 \quad \mbox{where} \, \; \theta \, \equiv \,  \alpha - \beta,
\label{thetadef}
\end{equation}
 will be the mixing angle in the 
neutral Higgs sector. Assuming the SM field $h^0$ to be close to $h$, means that
 $sin \theta$ is close to one.

In the Higgs basis the extended
 Yukawa interactions  may (in matrix notation) then be written
\begin{equation}
\, - \, \frac{v}{\sqrt{2}}{\cal L}_Y  \, =   \, 
\overline{Q_L}^0 \left( M_d^0 \, H_1 \, + \,  N_d^0 \, H_2 \right)
\, d_R^0  \, + \,  \overline{Q_L}^0 \,\left(M_u^0 \,
\tilde{H}_1 \, + \,  N_u^0 \,
\tilde{H}_2 \right) \, u_R^0 \ + \; h.c. \; ,
\label{NeutYuk}
\end{equation}
where 
\begin{equation}
M_d^0 = (c_\beta \Gamma_1 +
 e^{i \, \xi} s_\beta \Gamma_2) \frac{v \, e^{i \, \xi_1}}{\sqrt{2}} \; \; , \; 
N_d^0 = (s_\beta \Gamma_1 -
 e^{i \, \xi} c_\beta \Gamma_2) \frac{v \, e^{i \, \xi_1}}{\sqrt{2}} \; \, ,
\label{MNd}
\end{equation}
where $\xi = \xi_2 - \xi_1$ and for the $u$-quark case :
\begin{equation}
M_u^0 = (c_\beta \Delta_1 + 
e^{-i \, \xi} s_\beta \Delta_2) \frac{v \, e^{-i \, \xi_1}}{\sqrt{2}} \ \; \, , \; 
N_u^0 = (s_\beta \Delta_1 - 
e^{-i \, \xi} c_\beta \Delta_2) \frac{v \, e^{-i \, \xi_1}}{\sqrt{2}} \; \, .
\label{MNu}
\end{equation}
Now one transforms the mass matrices $M_{d,u}^0$ to diagonal form with matrices
$U_{R,L}^{d,u}$:
\begin{equation}
M_d \, = \, (U_L^d)^\dagger M_d^0 U_R^d \, = diag(m_d,m_s,m_b) \; \, , 
N_d \, = \, (U_L^d)^\dagger N_d^0 U_R^d \; ,
 \; d_{R,L} \, =  \, (U_{R,L}^d)^\dagger  \, d_{R,L}^0 \; ,
\label{MdNdiag}
\end{equation}
and similarly 
\begin{equation}
M_u \, = \, (U_L^u)^\dagger M_u^0 U_R^u \, = diag(m_u,m_c,m_t) \; \, , 
N_u \, = \, (U_L^u)^\dagger N_u^0 U_R^u \, ,
 \; u_{R,L} \, =  \, (U_{R,L}^u)^\dagger \, u_{R,L}^0 \; \, ,
\label{MuNdiag}
\end{equation}
for the $u$-quark case.

The total {\it neutral} Yukawa interactions for $d$-type quarks 
 may now be written in terms of physical quantities as \cite{Alves:2018kjr}:
\begin{equation}
\, - \, v {\cal L}_Y^{(d,n)}  \, =   \, 
\overline{d_L} \left( v + i G^0  - c_\theta \, H + s_\theta \, h
\right) M_d \, d_R  \, 
\, + \, \overline{d_L} \left( s_\theta \, H + c_\theta \, h + i A \right)
 \, N_d \, d_R
+ h.c
\label{NeutYukd}
\end{equation}
and for the $u$-quark case the corresponding interactions similarly
\begin{equation}
\, - \, v {\cal L}_Y^{(u,n)}  \, =   \, 
\overline{u_L} \left( v -  i G^0  - c_\theta H + s_\theta h
\right) M_u \, u_R  \, 
\, + \, \overline{u_L} \left(s_\theta H + c_\theta h - i A \right) \, N_u \, u_R
+ h.c
\label{NeutYuku}
\end{equation}

For the charged interactions one obtains
\begin{equation}
\, - \, \frac{v}{\sqrt{2}} {\cal L}_Y^{(charged)}  \, =   \, 
\overline{u_L}V_{CKM} \left(  G^+ M_d \, + \, H^+ N_d \right) d_R  \, 
\, - \, \overline{d_L} V_{CKM}^\dagger \left(G^- \, M_u +  H^- \, N_u \,\right) u_R
+ h.c
\label{NeutYukC}
\end{equation}

While the mass matrices $M_d$ and $M_u$ are now flavor diagonal, the matrices 
 $N_d$ and $N_u$ are in general flavor non-diagonal and CP-violating. 
These may give contributions  to  the $Y_R$'s in eq. (\ref{FCH-coupl}).
In the eqs. 
(\ref{NeutYukd}) and  (\ref{NeutYuku}) one observes that there will be
flavor changes for the lightest Higgs $h$ proportional to the non-diagonal
matrices $N_d$ and $N_u$. 

 In many papers on 2HDM, one assumes for
 instance (an) extra 
discrete  symmetry(-ies) to simplyfy the theory. 
In \cite{Alves:2018kjr} possible restrictions on 
(\ref{YukawaGendD}) and (\ref{YukawaGenu}) are discussed. 
Here I stick to the
 general case.

Further, I consider  how the 6-dimensional interaction in eq (\ref{FCNClad})
is obtained in 2HDMs.
One way   might be to consider the part of the Higgs potential
containing a product of four  $\Phi_1$ or $\Phi_2$
Higgs fields
 (see for example \cite{Grzadkowski:2016szj}): 
\begin{eqnarray}
V_{2HDM}^{4 \Phi}  \, =   \, \frac{1}{2} \, \lambda_1 (\Phi_1^\dagger \Phi_1)^2 \,
+ \, \frac{1}{2} \, \lambda_2 (\Phi_2^\dagger \Phi_2)^2 \, + \,
 \lambda_3 (\Phi_1^\dagger \Phi_1)(\Phi_2^\dagger \Phi_2) \, + \,
 \lambda_4 (\Phi_1^\dagger \Phi_2)(\Phi_2^\dagger \Phi_1) 
\nonumber \\
 \, + \, \frac{1}{2} \,\left[ \lambda_5 (\Phi_1^\dagger \Phi_2)^2 \, 
+ h.c.\right]  \, +  \, \ \left( \left[\lambda_6 (\Phi_1^\dagger \Phi_1) \, 
  + \lambda_7 (\Phi_2^\dagger \Phi_2) \right] (\Phi_1^\dagger \Phi_2) 
\, + \, h.c \right)
\label{HiggsPot}
\end{eqnarray}
Such  potentials may contribute to $Y_R$'s in (\ref{FCH-coupl}).
In (\ref{FCNClad}) $\phi$ is the SM Higgs which one within 2HDMs identify with 
$H_1$ in (\ref{SMHiggs}). Inserting (\ref{Higgsbase}) in 
 (\ref{HiggsPot}),  the Higgs potential (\ref{HiggsPot}),  will contain  
several terms of the type $(H_1^\dagger H_1)(H_2^\dagger H_1)$.
The coefficient for the sum of  such terms is  
\begin{eqnarray}
C_\lambda \, = \, \lambda_1 \, s_\beta c_\beta^3 
\, + \, \lambda_2 \, c_\beta s_\beta^3 \, + \, 
(\lambda_3 \, + \, \lambda_4) \, s_\beta c_\beta(s_\beta^2 -c_\beta^2) 
\, + \,   \, \lambda_5 \, e^{-2 i \xi} s_\beta^3 c_\beta \, - 
\, \lambda_5^* \, e^{2 i \xi} c_\beta^3 s_\beta \nonumber \\
\, + \,
2 \, \lambda_6 \,c_\beta^2 s_\beta^2 e^{-i \xi}  
 \, + \, \lambda_6^* (s_\beta^2 c_\beta^2 \, - c_\beta^4) e^{i \xi} \,
+ \, \lambda_7 \, (s_\beta^4  \, - \, s_\beta^2 c_\beta^2) e^{- i \xi} \, - \,
2 \, \lambda_7^* \, c_\beta^2 s_\beta^2 e^{i \xi} \; . 
\label{lambdas}
\end{eqnarray}
 Now the 
field $H_2^\dagger$  at space-time $z_2$ in such expressions might be 
contracted with the field $H_2$ at space-time $z_1$ in (\ref{NeutYuk}).
This  makes the field contraction (${\cal C}$)  :
\begin{equation}
{\cal C}\left(H_2(z_1) (H_2(z_2))^\dagger \right) \; = \;
D_{H_2}(z_1  \, - \, z_2) \, .
\label{Contr}
\end{equation}
Then one obtains an effective  6-dimensional interaction
\begin{eqnarray}
\, - \, \frac{v}{\sqrt{2}} \,  {\cal L}_6^{(D)}  \, &=&   \, 
(\overline{(Q_L(z_1))}^0) \left[ M_d \, H_1(z_1) \, \right. \nonumber \\
&+& \,
\left.
 C_\lambda  \, N_d  \, D_{H_2}(z_1 \, - \, z_2) \, H_1(z_2) 
((H(z_2)_1)^\dagger \, H_1(z_2)) \right] \, (d_R)_j^0(z_1)    + \; h.c. \; .
\label{YukawaGend}
\end{eqnarray}

So far, this is not a local operator. The propagator $D_{H_2}(z_1 \, - \, z_2)$
contains a propagator $D_h$ for the light neutral  Higgs, and a part
$D_H$ for the heavy neutral Higgs $H$. 
However, if the SM Higgs is close to the light Higgs $h$, then 
$s_\theta$ is close to one, and thereby the scalar $R_0$ is close to 
the heavy Higgs $H$.
The part containing the heavy Higgs $H$ interaction is then 
for $M_H^2 \gg M_h^2$ : 
\begin{equation}
D_{H}(z_1 \, - \, z_2) \, \simeq  \, -  \,
 \frac{\delta^{(4)}(z_1 - z_2)}{M_H^2} \, (sin \theta)^2 \; \, ,
\label{localH}
\end{equation}
making the $H$-part of the interaction local in this limit,
and giving the following non-diagonal contribution to $Y_R$ in (\ref{FCH-coupl})
and (\ref{FC-matr})  for $i \neq j$:
\begin{equation}
\left((Y^{(D)}_R)_{ij}\right)_H \, = \, - \, \left((\epsilon^{(D)}_R)_{ij}\right)_H
\, = \,  \frac{v^2 (\tilde{\lambda}_{ij})_H}{\sqrt{2} \Lambda_{NP}^2}  \, \simeq \, 
- \, C_\lambda (N_d)_{ij} \frac{v}{\sqrt{2} M_H^2}  \, (sin \theta)^2
 \; \, .
\label{FC-matrH}
\end{equation}

The non-local $h$-part $D_{h}$ would be a term corresponding to a 
 higher order diagram.
This term  is 
 shortly discussed at the end of section V.

\section{nEDM generated from a Flavor Changing Higgs coupling}

In this section  I give a short summary of the results from the
 previous paper \cite{Eeg:2016fsy}. The reason being that 
the diagrams calculated in that paper
are also relevant in  2HDM's.

\begin{center}
\begin{figure}[htbp]
\scalebox{0.70}{\includegraphics{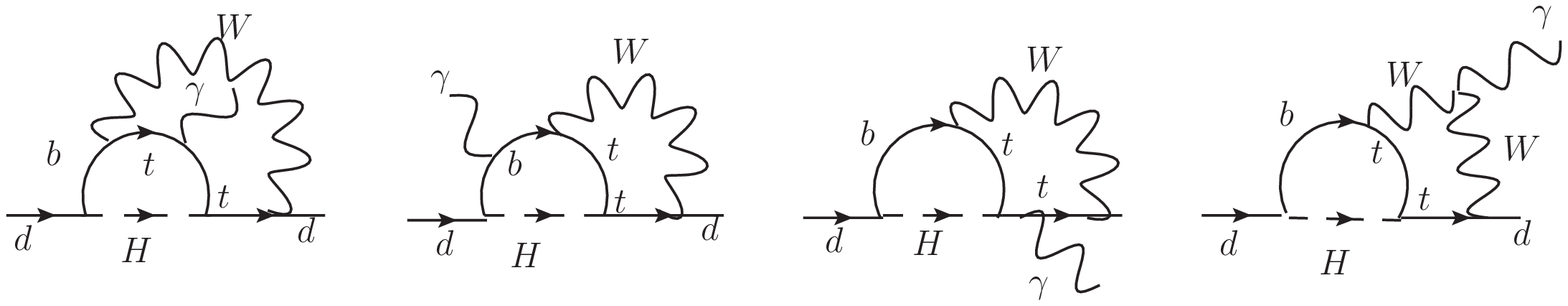}}
\caption{The first class of diagrams contain the FCH coupling and the big 
Higgs-top coupling proportional to the top mass $m_t$. 
For the first three diagrams, there are also corresponding diagrams 
where the $W$-boson is replaced by an Goldstone Higgs-boson within
 Feynman gauge. In this figure $H$ may denote the SM Higgs $h^0$. 
 Further, within the 2HDMs  $H$ may denote 
the lightest
  neutral  Higgs boson $h$,   
the heavier neutral Higgs boson $H$, or the neutral pseudoscalar Higgs $A$.
 }
\label{FCHNew2loopg}
\end{figure}
 \end{center}

\begin{center}
\begin{figure}[htbp]
\scalebox{0.6}{\includegraphics{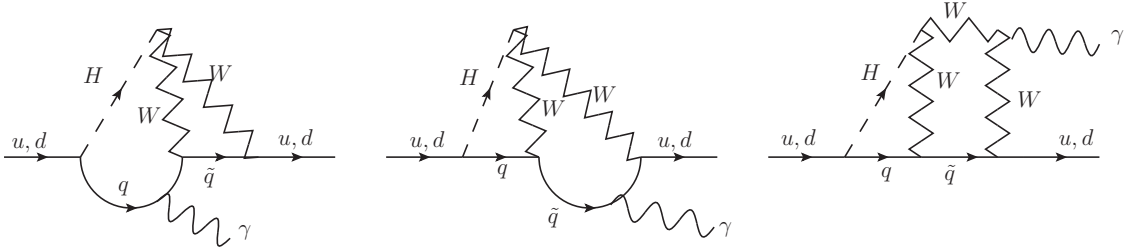}}
\caption{A class of diagrams containing the FCH coupling and the big 
$HWW$ coupling proportional to $M_W$. Additional graphs with the  W replaced
 by an unphysical (Goldstone) Higgs within Feynman gauge has to be added. In  
this figure $H$ may denote the SM Higgs $h^0$, or within the 2HDMs the lightest
  neutral Higgs $h$, 
the heavier neutral Higgs $H$, or the neutral pseudoscalar Higgs $A$.
 }
\label{2LoopWWHgam}
\end{figure}
 \end{center}

In \cite{Eeg:2016fsy},
two classes of diagrams for EDMs of light quarks, shown
 in Fig. \ref{FCHNew2loopg} and
 Fig. \ref{2LoopWWHgam}, were considered.
These  diagrams are  obtained from the flavor non-diagonal interaction in 
(\ref{FCH-coupl}), completed by  SM interactions.
But these diagrams  also give  contributions within 2HDM with flavor change,
as explained in the next section. 

As explained in \cite{Eeg:2016fsy},
  the $u$-quark dipole 
moment $d_u$ due to diagrams in Fig.\ref{FCHNew2loopg} are suppressed,
 and therefore neglected. Thus,  the $d$-quark dipole moment 
 contributions dominate, and other contributions are 
neglected.
Summing all contributions from diagrams in Fig. \ref{FCHNew2loopg} 
and Fig. \ref{2LoopWWHgam},  I obtained the
 dominanting contribution in the bare case (before QCD corrections)
\cite{Eeg:2016fsy}:
\begin{eqnarray}
 (\frac{d_d}{e})_{Tot} \, =  \, F_2 \, C_E(\mu_\Lambda) \,  \,
 Im[Y_R(d \rightarrow b) \, V_{td}^* \, V_{tb}].
\label{eq:AmpTot}
\end{eqnarray}
where  the constant $F_2$ sets the overall scale of the EDMs obtained from the
 two loop diagrams:
\begin{equation}
F_2 \, = \,  \frac{g_W^3}{M_W \, \sqrt{2}} \left(\frac{1}{16 \pi^2}\right)^2
\, = \, \frac{2 M_W^2}{v^3} \left(\frac{1}{16 \pi^2}\right)^2 \,
 \simeq \, 6.94 \times 10^{-22} \, \mbox{cm} \; ,
\label{F-quantity}
\end{equation}
where $v \, = $ 246 GeV is the electroweak symmetry breaking scale,
 and where I have used the conversion 
relation $1/(200 \mbox{MeV}) = 10^{-13}$ cm.

Further,
\begin{equation}
C_E(\mu_\Lambda) \, = \, 
 \left( \left[\frac{2 u_t}{3} p_1(u_t)  + \frac{25}{12} p_2(u_t) \right] 
\, C_\Lambda  \, + f_{Fin}\right)  \, .
\label{D-amp}
\end{equation}
Here $p_1$ and $p_2$ are one loop functions for finite 
subloops \cite{Eeg:2016fsy} :
\begin{equation}
p_1(u) \, \equiv \, \frac{u}{(u-1)} \left(1 - \frac{ln(u)}{u-1} \right) 
\; \; \,  ; \; \; \mbox{and} \; \; \; 
p_2(u) \, \equiv \,
 \frac{u}{(u-1)} \left( \frac{u \cdot ln(u)}{u-1} \, - 1 \right) \; . 
\label{p-func}
\end{equation}
Using standard values for the masses of $W$ and  $t$ ,
 one finds  numerically 
\begin{equation}
 u_t \equiv \left(\frac{m_t}{M_W}\right)^2 \simeq 4.65 \; \; ; \; \,
 p_1(u_t) = 0,737 \; \, , \; \,  p_2(u_t) = 1.219 \, .
\label{ut}
\end{equation} 

The UV divergence is
 parametrized through the quantity
\begin{equation} 
C_\Lambda \;\equiv  \; ln(\frac{\Lambda^2}{M_W^2}) \, + \, \frac{1}{2} \; \, ,
\label{P-CLambdiv}
\end{equation}
where $\Lambda$ is the UV cut-off. Numerically,  $C_\Lambda $ is $\sim 5.5$
to 9.4 for $\Lambda \sim $ 1 to 7 TeV . 
The quantity $f_{Fin} \simeq -7.7$, is the sum  of the diagrams not containing 
divergent parts , and also 
 the finite parts of diagrams containing a divergence.
The divergence appears in the $d \rightarrow t \, W $ subloop in 
some of the diagrams. 

 The $V$'s are the Cabibbo-Kobayashi-;askawa (CKM) matrix 
elements in the standard notation.
We note that because  $ V_{td}^* \, V_{tb}$ is complex, there will be an
 EDM even if $Y_R(d \rightarrow b)$ is real  !

 Using  the absolute value of  $V_{td}^* \, V_{tb}$ from 
 \cite{Tanabashi:2018oca}, one may write my result for the nEDM in the 
following way, as shown in \cite{Eeg:2016fsy}:
\begin{equation}
d_n/e   \simeq  N(\Lambda) \times 
\left\{ \frac{|Y_R(b \rightarrow d)|}{|Y_R(b \rightarrow d)|_{Bound}} \cdot
\mbox{Im} \left[\frac{Y_R(d \rightarrow b)}{|Y_R(b \rightarrow d)|}
 \cdot \frac{V_{td}^* \, V_{tb}}{|V_{td}^* \, V_{tb}|}\right] \right\}
 \times 10 ^{-26} \,  \mbox{cm} \; ,
\label{dn-f}
\end{equation}
where I have scaled the result with the bound in 
 \cite{Blankenburg:2012ex,Harnik:2012pb} obtained
 from $B_d - \overline{B_d}$-mixing:
 \begin{equation}
| Y_R(d \rightarrow b)| \;  \le \; 1.5 \times 10^{-4} \;  \equiv \; | Y_R(d \rightarrow b)|_{Bound} \, .
\label{YRBound}
\end{equation}
The function $N(\Lambda)$ is defined by the relation
\begin{equation}
 \rho_d \, F_2 \, |Y_R(d \rightarrow b)|_{Bound}  |V_{td}^* \, V_{tb}| 
\, C_E(\mu_\Lambda) = \, N(\Lambda)  \, \times 10 ^{-26} \,  \mbox{cm} \, ,
\label{NLambda}
\end{equation}
where $\rho_d \simeq 0.74$ is the contribution from the $d$-quark EDM $d_d$ to
 the neutron EDM $d_n$ within lattice calculations \cite{Bhattacharya:2015esa,Bhattacharya:2015wna,Yamanaka:2018uud}.

\begin{center}
\begin{figure}[htbp]
\scalebox{0.55}{\includegraphics{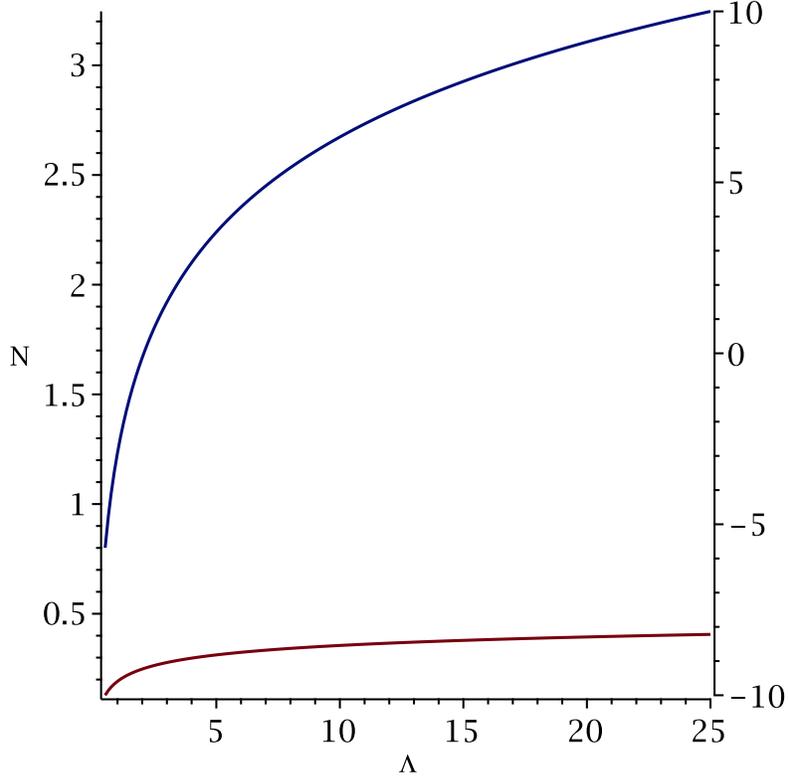}}
\caption{The quantity $N \, = \, N(\Lambda)$, in units $10^{-26}$ cm, 
 as a function  of cut-off $\Lambda$ (in TeV). The blue (upper) curve is for the bare case, and the red (lower) curve is for the case when the suppressing 
QCD corrections are included.}
\label{nEDM-numerics}
\end{figure}
 \end{center}

 $N(\Lambda)$
 is plotted as a
function of $\Lambda$ in Fig. \ref{nEDM-numerics} for the bare case
(at the renormalization scale $\mu = \mu_\Lambda$, blue curve) and
 with QCD corrections (at the hadronic scale $\mu = \mu_h \, \simeq$ 1 GeV,
 red curve, as explained in \cite{Eeg:2016fsy})

 Now, the {\it maximal value} of 
the parenthesis $\{...\}$  in (\ref{dn-f}) is $= 1$. 
Thus, {\it if} the bound for $Y_R(d \rightarrow b)$ in (\ref{YRBound}) is
 saturated,
 the plot for the function $N_\Lambda$  in Fig. \ref{nEDM-numerics}
shows that when the cut-off $\Lambda$ is stretched up to 20 TeV, the bound for 
nEDM in (\ref{dn-bound}) is reached in the bare case, while the perturbative
 QCD-suppression tells  \cite{Eeg:2016fsy}
 that the value of the nEDM can at maximum be 
 at most  $0.4 \times 10^{-26} \, e$ cm.
 for $\Lambda$ up to 20 TeV. If the
 bound for $|Y_R(d \rightarrow b)|$ is reduced, and also $\Lambda$ is reduced,
 my value for nEDM will be accordingly smaller.

\section{The nEDM in the 2HDMs}

All  diagrams in Fig. \ref{FCHNew2loopg} 
and Fig. \ref{2LoopWWHgam} will also contribute within 2HDM 
without restrictions as given in
(\ref{NeutYukd}) and  (\ref{NeutYuku}), and one obtains diagrams both
 with light $h$- and heavy $H$- and $A$-exchanges. Exchange of 
the pseudoscalar $A$ 
(as defined in eq. (\ref{SMHiggs}))
does not contribute to the order I work, because it does not couple to 
the mass matrices $M_u$ and $M_d$ in
(\ref{NeutYukd}) and (\ref{NeutYuku}). As also is seen from these equations, 
$A$ couples only to the matrices $N_{u,d}$. Exchanges of $A$ will be 
treated  explicitly later in this section, and are shown to be suppressed. 
For  contributions with exchange of $h$ and $H$,
 the amplitudes are equal, 
 but have  
 opposite signs according to the eqs. (\ref{NeutYukd}) and  (\ref{NeutYuku}).
 Explicitly, for $X=h$ and $X=H$ exchanges I find the effective contribution
\begin{equation}
\left[Y_R(d \rightarrow b)\right]^{eff}_N \; = \; \frac{1}{v} 
\left( N_d \right)_{bd} \, cos\theta \, sin\theta  \; ,
\label{YNcs}
\end{equation}
where 
$\theta$ is defined in (\ref{thetadef}) and  
$\left( N_d \right)_{bd}$ in (\ref{MNd}). Within 2HDMs, given by  
(\ref{NeutYukd}) and (\ref{NeutYuku}), 
eq. (\ref{YNcs}) may the be 
inserted in (\ref{dn-f}). Combining (\ref{YNcs}) and 
(\ref{YRBound}) gives then a restriction on flavor changing 2HDMs from
$B_d -\overline{B_d}$-mixing \cite{Harnik:2012pb}.

As shown in \cite{Eeg:2016fsy} some diagrams with exchange of only one neutral 
Higgs-boson  
have divergent parts. 
In the following I will for illustrative purposes 
consider in some detail  the case where a
 soft photon is emitted from a $W$-boson, as shown in the right panel of 
 Fig. \ref{FCHNew2loopg}. 
Then the result for this diagram is proportional to the two loop integral 
tensor (neglecting $m_b$ compared to $m_t$):
\begin{equation}
T_{\mu \nu}^W(X) \, = \, \int \int 
\frac{\dbar p \, \dbar r \;K_\mu \, p_\nu }{(r^2 - M_W^2)^2 (r^2 - m_t^2)
((r+p)^2 - m_t^2)(p^2 - m_b^2)(p^2 - M_X^2)} \; \, ,
\, \;   
\label{loopint}
\end{equation}
where $M_X$ is the mass of either the light or heavy Higgs,
 {\it i.e} $X=h,H$, later also $A$.   
Moreover, $K_\mu \, = \, K^R_\mu \, = \, r_\mu$ when a 
Higgs is coupling to the top-quark with a 
right-handed coupling, and $K_\mu \, = \, K^L_\mu \, = \, (r+p)_\mu $
 when this coupling is left-handed.
In the latter case,  
the integral over $\dbar p$ diverges. When $X=h$ and $X=H$ couples to the
 diagonal mass matrix part $(M_u)_{tt} = m_t$ one has 
$K_\mu \, \rightarrow  (K^R_\mu \, + \, K^L_\mu) \, = \, (2r+p)_\mu$

In the limit where $sin \theta$ is close to one,
 the $h$-part can be written as in \cite{Eeg:2016fsy} :
\begin{equation}
 T_{\mu \nu}^W(h) \, = \, \frac{g_{\mu \nu}}{4 \, m_t^2} \, 
\left(\frac{1}{16\pi^2}\right)^2 \left(C_\Lambda \, \cdot  p_2(u_t) + 
t_{WFin}^L \, + \,t_{WFin}^N \right) \, ,
\label{loopint-h}
\end{equation}
where $C_\Lambda$ is given by (\ref{P-CLambdiv}) and 
where  $p_2(u)$ is defined in (\ref{p-func}) and $u_t$ is the mass 
ratio in (\ref{ut}). Furthermore,  
\begin{equation}
t_{WFin}^L \simeq -2.8 \quad , \; {\mbox and} \;  \,t_{WFin}^N \, \simeq \,
-1.1 \; .
\label{finite-terms}
\end{equation}
Here $t_{WFin}^L$ is the finite term following  the logarithmic divergence, 
and $t_{WFin}^N$ is a completely finite term,- as explained in the Appendix.
For other diagrams there are similar expressions as (\ref{finite-terms}), 
but with other numbers. Completely finite diagrams have only a
 term similar to $t_{WFin}^N$, for instance the two first (from left) 
diagrams in Fig. 1. 

One should remember that the loop amplitude in (\ref{loopint-h}) is 
multiplied by a factor $g_W m_t/M_W$ from the SM Higgs coupling to the
 top quark.
There will also be an extra factor $m_t$ from the mass part of  
 one of the top quark propagators. Thus the  factor $1/m_t^2$ at the right
 hand side of (\ref{loopint-h}) will  cancel and the quantity $C_E$  in 
  (\ref{D-amp}) is dimensionless. The factor $1/M_W$ from the Higgs to 
top quark coupling goes into $F_2$ in (\ref{F-quantity}).

The individual loop integrals for $X=h$ or $X=H$ alone  have divergent
 parts. But  within 2HDM, one  observes from eqs. (\ref{HiggMix}), 
(\ref{NeutYukd}) and (\ref{NeutYuku})
that the terms with exchange of $h$ and $H$ will have opposite signs 
  due to the
Cabibbo-like mixing of $h$ and $H$, and one ontains a  cancellation
 of  divergences due to exchanges of these two bosons.
Thus , I use
\begin{equation}
\frac{1}{(p^2 - M_h^2)} \, - \, \frac{1}{(p^2 - M_H)} \; = \;
\frac{(M_h^2 -M_H^2)}{(p^2 - M_h^2)(p^2 - M_H^2)} \; ,
\label{PropDiff}
\end{equation}
and obtain the total tensor for exchanges of both $h$ and $H$:
\begin{equation}
\Delta T_{\mu \nu}^W \, = \, T_{\mu \nu}^W(h) \, - \, T_{\mu \nu}^W(H) \, = \, 
(M_h^2 \, - \, M_H^2) \, S_{\mu \nu}^W \; \, ,
\label{DeltaT}
\end{equation}
where 
\begin{equation}
S_{\mu \nu}^W \, = \, \int \int 
\frac{\dbar p \, \dbar r \;(2r+p)_\mu \, p_\nu }{(r^2 - M_W^2)^2 (r^2 - m_t^2)
((r+p)^2 - m_t^2)(p^2 - m_b^2)(p^2 - M_h^2)(p^2 - M_H^2)} \; ,
\, \;   
\label{loopint-S}
\end{equation}
is finite.

As shown in the Appendix, this loop  integral contains logarithmic 
and dilogarithmic  functions of masses of the top quark, the $W$-boson and the 
neutral Higgs bosons $h,H$.
Inn the limit $M_H^2 \gg M_h^2$,
 I find that 
\begin{equation}
S_{\mu \nu}^W \; \sim \frac{ln(M_H^2)}{M_{SM}^2 M_H^2} \; ,
\label{Lead}
\end{equation}
where  $M_{SM}$ is either  $m_t$, $M_h$ and/or $M_W$.
 I have found the leading result replacing eq. (\ref{eq:AmpTot}) can
 be written: 
\begin{eqnarray}
 \, \Delta T_{\mu \nu}^W \, = \, 
\frac{g_{\mu \nu}}{4 m_t^2} \, 
\left(\frac{1}{16\pi^2}\right)^2 \left(\widetilde{(C_{H}})^W \,
\cdot \, p_2(u_t) + 
t_{WFin}^L \, + \,t_{WFin}^N \right) \, + \, {\cal O}(M_{SM}^2/M_H^2) \, \, ,
\label{loopint-DT}
\end{eqnarray}
where corrections are of order $(M_{SM}/M_H)^2$.
Here one might expect that  the divergent term $C_\Lambda$ from
 (\ref{loopint-h}) is replaced by a finite term where the cut-off $\Lambda$
is replaced by just the heavy neutral Higgs mass $M_H$.
This is true to leading order, but it turns out 
that the  
$(\widetilde{C_{H}})^W$
is a bit more complicated, 
 as shown in the Appendix:
\begin{equation}
(\widetilde{C_{H}})^W \; = \; 
ln \left[\left(\frac{(\widetilde{M_{H}})_W}{M_W}\right)^2 \right] \, 
+\,  \frac{1}{2} \; ,
\label{CH}
\end{equation}(up to corrections  of order
 $(M_{SM}/M_H)^2$ as mentioned above)
where
\begin{equation} 
(\widetilde{M_{H}})_W \; = \, M_H \, e^{\alpha_W} \; \; , \;\alpha_W \, \equiv \,
 \frac{(ln u_t)^2}{4(1-1/u_t -(ln u_t))} \; \, , \;   e^{\alpha_W} \simeq 0.45 \; ,
\label{CutoffConnection}
\end{equation}
and where $u_t$ is given in (\ref{ut}).
The term $t_{WFin}^L$  in (\ref{loopint-DT})
 is, up to $(M_{SM}/M_H)^2$,  the same $t_{WFin}^L$  as in (\ref{loopint-h}).

 It is easy to see that 
the $t_{WFin}^L$'s are the same 
if one uses the mathematical trick given in (\ref{trick}) in the Appendix. 
The term $t_{WFin}^N$ is trivially the same (up to corrections of order 
$(M_{SM}/M_H)^2$).
For other diagrams, where the soft photon is emitted by a quark $q = b,t$, 
say, the factor $e^{\alpha_W}$ will be replaced by a similar 
factor $e^{\alpha_q}$  of the same order of magnitude.
Now the result given by (\ref{YNcs}), and
 (\ref{loopint-DT})-(\ref{CutoffConnection}) can be    completed with 
similar expressions for the rest of digrams in Figs. 1 and 2.
Then the final resiult will be as in \cite{Eeg:2016fsy}, {\it i.e} as in  eq. 
(\ref{dn-f}) and Fig. 3, with the cut-off $\Lambda$ replaced by a mass 
$\widetilde{M_H}$ of order $M_H$ (depending on the various 
$\alpha$'s similar to 
$\alpha_W$ in (\ref{CutoffConnection}) ).

Up to now I have considered contributions where $X=h$ and $X=H$ are coupled to
the diagonal mass matrix $M_u \rightarrow m_t$. Now I will consider the
 contributions where  the neutral Higgses $h,H,A$ couple to the
 diagonal $tt$ element of $N_u$.  Due to the mixing of the scalars 
 $h$ and $H$ the exchanges of these are given, as seen from (\ref{NeutYukd}) and
(\ref{NeutYuku}),  by the  propagator terms
\begin{equation}
\frac{(sin \theta)^2}{(p^2 - M_H^2)} \, - \, 
\frac{(cos \theta)^2}{(p^2 - M_h)} \; = \;
\frac{1}{(p^2 - M_H^2)} \, - 
\frac{(cos \theta)^2(M_H^2 -M_h^2)}{(p^2 - M_h^2)(p^2 - M_H^2)} \; ,
\label{PropDiffTh}
\end{equation}
The last term on the right-hand side will give finite and very small
 terms because $cos  \theta$ is small. Such terms are then neglected.
The first term on the right-hand side will a priori give a divergent 
term if $H$ has a left-handed coupling to $(N_u)_{tt}$,
({\it i.e.} $K_\mu = K^L_\mu = (r+p)_\mu$ and $X=H$ in (\ref{loopint})).
 However, because of the imaginary
 coupling of $A$, the similar exange of $A$ will also be divergent and  
cancels the divergent term from $H$. The cancellation is exact in the limit 
$M_A \rightarrow M_H$, and in general  there is a finite leftover. 
It is important to note that such (partial) cancellations occur also
 {\it before}
eventual explicite resrictions (symmetry requirements) 
are assumed for the  2HDMs.
When $H$ and $A$ couples righthanded with $(N_u)_{tt}$,
they have the same sign and are finite, and they are  equal in the 
limit $M_A \rightarrow M_H$. In this case one finds a tensor with 
the leading behaviour
  \begin{equation}
T_{\mu \nu}^{WR} \; \simeq  - (\frac{1}{16 \pi^2})^2 \frac{g_{\mu \nu}}{32 M^2}
 ln(\frac{M^2}{M_{SM}^2}) \, ,
\label{LeadTA}
\end{equation}
where $M =M_H$ or $M=M_A$, and $M_{SM}$ is of order $M_W$ or $m_t$.
Thus, the contributions from $(N_u)_{tt}$ are suppressed by
$(M_{SM}/M))^2$ compared to the terms with coupling to the mass matrix
  $M_u$ as stated in the beginning of this section.
Then, my result has, before QCD corrections, the general structure
\begin{eqnarray}
 (\frac{d_n}{e})   \, \sim (\frac{1}{16\pi^2})^2 
\, \frac{g_W^3}{M_{SM}} \left[ln(\frac{M_H^2}{M_{SM}^2}) \, + \, NL \right] \,
\cdot  \, Im[Y_R(d \rightarrow b) \, V_{td}^* \, V_{tb}] \, ,
\label{eq:AmpGen}
\end{eqnarray}
where $NL$ is a dimensionless non-logarithmic, non-leading term, 
depending on $m_t$, $M_W$, and the light higgs mass $M_h$. 
$NL$ also contains the $(M_{SM}/M)^2$ corrections.

Concerning the six-dimensional interaction in (\ref{FCNClad}), it will 
for the local part also be proportional to $N_d$ as 
shown in (\ref{FC-matrH}), and
 also suppressed by $(v/M_H)^2$. The   non-local part given by
 exchange of the light $h$-boson will be of one order higher. The corresonding 
loop diagram
is proportional to
\begin{equation}
T_{\mu \nu}^{W6} \, = \,S_{\mu \nu}^W(M_H=M_h) \; = \; g_{\mu \nu} 
 \frac{1}{(16\pi^2)^2} \frac{1}{4 M_W^4} \times 0.015 \, ,  
\label{T6}
\end{equation}
where standard numerical values for masses for SM particles has been used. 
The contribution from this diagram should also be multiplied by
$(cos \theta)^2  \, v^2 $, and $C_\lambda$ in (\ref{lambdas}),
 and will be  small (because $cos \theta$ is small
when $h$ is close to $h^0$) .

\begin{center}
\begin{figure}[htbp]
\scalebox{0.50}{\includegraphics{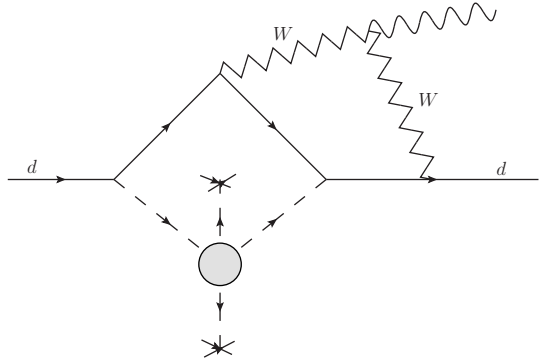}}
\caption{Diagram generated in the 2HDMs. The blob denotes the 
interaction in (\ref{HiggsPot}).  The dashed lines are Higgses and
 the crosses in the end of two of these denotes the Higgs VEV.}
\label{La67loop}
\end{figure} 
 \end{center}

\section{Discussions and Conclusions}

In previous papers \cite{Blankenburg:2012ex,Harnik:2012pb} based on
 the effective theory for flavor changing Higgs (FCH) couplings, one loop
 diagrams for the neutron EDM were considered. There is a one loop diagram
 for an EDM of the $u$-quark with Higgs and the $t$-quark in the loop
 which is proportional to the $t$-quark mass and the product of the 
FCH couplings  $Y_R(u \rightarrow t)$ and  $Y_R(t \rightarrow u)$.
The absolute values of these FCH couuplings are not very restricted, according
to \cite{Harnik:2012pb}:
\begin{equation}
\sqrt{|Y_R(u \rightarrow t)|^2 \, + \,|Y_R(t \rightarrow u)|^2} \leq 0.3 \, .
\label{B1}
\end{equation}
However, from the nEDM based on this one loop diagram one obtains
the bound
\begin{equation}
Im \left[Y_R(u \rightarrow t) \times Y_R(t \rightarrow u)\right] 
\leq  4.3 \times 10^{-7} \, .
\label{B2}
\end{equation}
 There is a one loop diagram
 for an EDM of the $d$-quark with Higgs and the $b$-quark in the loop 
proportional to  the $b$-quark mass and the product of FCH couplings
 $Y_R(d \rightarrow b)$ and  $Y_R(b \rightarrow d)$.
Bound on the coupling $Y_R(d \rightarrow b)$ is given in 
(\ref{YRBound}), and one has also \cite{Harnik:2012pb}
\begin{equation}
Im \left[Y_R(d \rightarrow b) \times Y_R(b \rightarrow d)\right] 
\leq  6.4 \times 10^{-8} \, .
\label{B3}
\end{equation}
Then the  $d$-quark EDM with a $b$-quark one loop diagram could at most
 give a contribution to the nEDM of about 
$1.3 \times 10^{-26} \, e$ cm before QCD corrections are taken into account.
But QCD corrections will suppress this result further by a
 factor of order $10^{-1}$ (see \cite{Eeg:2016fsy} and references therein).

In my previous paper  \cite{Eeg:2016fsy} I presented calculations of 
two loop diagrams which depended on a flavor changing coupling.
Such two-loop contributions are suppressed with the flavor changing coupling
 to {\it first order only}, instead of the second order suppression for 
one loop diagrams. 
A price for going to two loops is in this case a suppression factor 
$g_W^2/(16 \pi^2) \simeq 1.2 \times 10^{-3}$
 which is numerically 
bigger than $Y_R(d \rightarrow b)$, as seen in eq. (\ref{YRBound}).
Therefore the two loop diagrams of ref. \cite{Eeg:2016fsy}, are expected to
(more than) compete numerically with the corresponding one loop diagrams.  
Also it is important to note that in the two loop case considered here, 
 there will be a EDM different 
from zero even if  $Y_R(d \rightarrow b)$  is real, 
because it is combined with a CKM factor.

Going from the effective theory  of
 \cite{Blankenburg:2012ex,Harnik:2012pb,Eeg:2016fsy} to the general 2HDM-model
in this paper, I have  shown that 
the result from \cite{Eeg:2016fsy} stays the same, up to corrections
 of order  $(M_{SM}/M_{H,A})^2$,
when the flavor changing coupling 
 \cite{Blankenburg:2012ex,Harnik:2012pb,Eeg:2016fsy}, is replaced by
 the correponding expression in 2HDMs, as shown in eq. (\ref{YNcs}).
This equation then gives a bound for the expression 
$(N_d)_{bd} \cdot cos \theta \cdot sin \theta/v$ within 2HDM with flavor change.
Further, I have shown that
the divergences appearing for some 
EDM diagrams with flavor changing Higgs in my previous
 paper \cite{Eeg:2016fsy} are, as expected, 
removed when extending the analysis to
Two Higgs Doublet Models  allowing for flavor changes by neutral scalars. 
Namely, the divergences $\sim ln(\Lambda)$ in
\cite{Eeg:2016fsy} are replaced by  $ ln(\widetilde{M_H})$, where 
$\widetilde{M_H}$ is equal to $M_H$ times a function of the ratio $m_t/M_W$.
(see eq. (\ref{CutoffConnection})).
Finally,  I have demonstrated that exchanges of the 
pseudoscalar Higgs $A$ does {\it not contribute to the order I work}
 because they are suppressed by $(M_{SM}/M_A)^2$. I have also demonstrated
 that some potential divergences involving exchanges of $A$ are cancelled 
by similar divergences  from exchanges of the heavy scalar $H$.

It is also found that the six dimensional interaction in (\ref{FCNClad})
 used in  \cite{Eeg:2016fsy} will also be proportional to the  
non-diagonal $N_d$-matrix in (\ref{NeutYuk}), (\ref{MdNdiag}) and 
 (\ref{NeutYukd}). 
 The quantity $C_\lambda$ from the Higgs potential contains many unknowns.
But this term is suppressed by $(v/M_H)^2$, as seen from 
eq. (\ref{FC-matrH}).

There are also other calculations of the nEDM 
\cite{Crivellin:2013wna,Jung:2013hka}. In \cite{Crivellin:2013wna} EDM for 
flavor changing couplings are considered, but only at one loop.
The results, given by the matrices in eq. (75) of that paper are 
reasonable agreement with
\cite{Harnik:2012pb}, cfr. also the eqs.(\ref{B1}),(\ref{B2}),(\ref{B3}).

 In \cite{Jung:2013hka} a 2HDM is considered at two loop level
in terms of Barr-Zee diagrams, but  with no flavor change from neutral Higgses.
This contribution is suppressed by the mass ratio $m_f/M_W$, where $m_f$ is a
light quark mass $m_u$ or $m_d$.
Then,  if the 
 non-diagonal elements of the matrix $N_{d}$ are   of he same order as  the
diagonal ones, the result presented in this paper is bigger than in eq. (53)
of  \cite{Jung:2013hka}.  On the other hand, if the non-diagonal element
$(N_d)_{bd}$ is very small, the result of \cite{Jung:2013hka} might be bigger.
And of course, if the non-diagonal elements of $N_d$ are
 restricted to be zero to avoid flavor changing neutral currents
 completely, then my result is zero. 

In general, the mechanism given by the diagrams in Fig. 1 will also 
work in some other theories with exchanges of scalars and the $W$-boson, 
for example for an EDM of the electron within leptoquark models 
\cite{Dorsner:2016wpm}.

To conclude,  when going from \cite{Eeg:2016fsy} to
the present study of 2HDMs with flavor change I have shown that:

\begin{itemize}

\item
 The result from \cite{Eeg:2016fsy} stays unchanged up to corrections of 
$(M_{SM}/M_{H,A})^2$.
The  logarithmic divergence $\sim ln(\Lambda/M_W)$ in \cite{Eeg:2016fsy}
 is replaced by 
 $ln(\widetilde{M_H}/M_W)$, where $\widetilde{M_H}$ is of order $M_H$.

\item

 The flavor changing coupling $Y_R(d \rightarrow b)$ in  
\cite{Blankenburg:2012ex,Harnik:2012pb,Eeg:2016fsy}
is found to be replaced by $(N_d)_{bd}  \cdot cos \theta \cdot sin \theta/v$
 in  2HDMs with flavor changeing neutral Higgses. 
 I have also identified an example of the six-dimensional term 
 in eq. (\ref{FCNClad}) which was a starting point in
  \cite{Harnik:2012pb,Dorsner:2015mja,Eeg:2016fsy}.

\item
 There is a cancellation between divergent terms with $A$- and $H$-exchanges.

\item
There is a  suppression  $(M_{SM}/M_{H,A})^2$ 
of finite terms due to exchange of $A$ and $H$-terms not 
coupled to the top mass.

\end{itemize}
 
\section{ Appendix. Details for the loop integrals.}

If the soft photon is emitted from the $W$-boson
 as in the left diagram
  in Fig. \ref{FCHNew2loopg},
then  the left sub-loop 
containing the Higgs boson 
is logarithmically divergent. The result of the divergent part
of (\ref{loopint}) 
can be written
\begin{eqnarray}
T_{\mu \nu}(h) \, = \, \frac{g_{\mu \nu}}{4} \, 2! \,
  \,  \int_0^1 dx \int_0^{(1-x)} dy 
 \int \frac{\dbar r }{(r^2 - M_W^2)^2 (r^2 - m_t^2)} \,  
\left(I_2(R) \, + R \, \cdot I_3(R)\right) \; ,
\label{Divloop1}
\end{eqnarray}
where  the quantity $R$ depends on the squareed loop momentum $r^2$
For  $n=2,3$ :
\begin{eqnarray}
I_n(R) \, = \,  \int \frac{\dbar p }{(p^2 - R)^n} \; .
\label{Divloop2}
\end{eqnarray}
Then for cut-off regularization :
\begin{eqnarray}
\left(I_2(R) \, + R \, \cdot I_3(R)\right) \, = \, 
\, \frac{i}{16 \pi^2}
\, \;   \left( ln(\Lambda^2/R) - \frac{3}{2}  \right) \; \, ,
\label{DivSubloop3}
\end{eqnarray}
where $\Lambda$ is the cut-off, and $x$ and $y$ are Feynman parameters, and
\begin{equation}
R \, \equiv\, B \, - \, x(1-x)r^2 \; ;  \quad 
 B \, \equiv \, m_b^2 \, + \, x (M_W^2 -m_b^2) \, + \, y (M_h^2 -m_b^2)
\; \, .
\label{Rvalue}
\end{equation}

One may split up 
\begin{equation}
\left( ln(\Lambda^2/R) - \frac{3}{2}  \right) \; = \; 
\left( ln(\Lambda^2/M_W^2) - \frac{3}{2}  \right) 
\, + ln(M_W^2/R) \, \; ,
\label{Logsplit}
\end{equation}
where the first term corresponds to $C_\Lambda$
 in (\ref{P-CLambdiv}), and 
   the $ln(M_W^2/R)$ term correspond to $t_{WFin}^L$.  
There is also  a finite term $t_{WFin}^N$ corresponding to 
an completely finite extra term $ \sim 1/R$ not shown in (\ref{Divloop1}).
We also note that the one loop integral
\begin{equation}
K_W \, = \,  \int \frac{\dbar r }{(r^2 - M_W^2)^2 (r^2 - m_t^2)} \; = 
\frac{- i}{16 \pi^2 m_t^2} p_2(u_t) \; , 
\label{Kloop}
\end{equation}
where $p_2(u)$ defined in (\ref{p-func}) is the proportionality 
factor for the divergent term $C_\Lambda$ in (\ref{P-CLambdiv}) .

Now I consider the finite loop integral in (\ref{loopint-S}) with
 both $h$ and $H$ included.
Doing Feynman parametrisation for the $\dbar p$-integration one obtains
\begin{equation}
S_{\mu \nu}^W \, = \, \, \frac{g_{\mu\nu}}{4} \; S^W   \; , \quad   
S^W \, = \, 2 ! \, \left( \frac{i}{16 \pi^2} \right)  \, 
 \int_0^1\frac{dx}{x(1-x)}\int_0^{(1-x)}dy \int_0^{(1-x-y)}dz \, J_Q^W \, ,
\label{loopintJ}
\end{equation}
plus terms suppressed by $1/M_H^2$. This integral is finite.
Note that the term $\sim 1/R$ mentioned just above  (\ref{Kloop}) 
 is  not included.
Here 
\begin{eqnarray}
&\,&J_Q^W \,  \equiv \, \int\frac{\dbar r}
{(r^2 - M_W^2)^2 (r^2 - m_t^2)(r^2 - Q)}  \nonumber \\
&=& \; 
\frac{i}{16\pi^2}\left[\frac{-1}{(m_t^2-M_W^2)(Q-M_W^2)}
+ \frac{m_t^2}{(m_t^2-M_W^2)^2}\left(\frac{ln(Q/M_W^2)}{(Q-M_W^2)} \, - \, 
\frac{ln(Q/m_t^2)}{(Q-m_t^2)}\right) \right] \; ,
\label{loopintQ}
 \end{eqnarray}
where 
\begin{equation}
Q \, = \, \frac{1}{x(1-x)} \left(m_b^2 + x (m_t^2 - m_b^2) \, + 
 y (M_h^2 - m_b^2) + z (M_H^2-m_b^2) \right) \; .
\label{Qval}
\end{equation}
Integrating over $z$ gives a suppression factor of order $1/M_H^2$. 
Changing variables, one obtains an integral over $Q$
with $dz \, = \, x(1-x) dQ/M_H^2$ :
\begin{eqnarray}
\int_0^{(1-x-y)} \, dz \, J_Q^W \,  = \,  
\frac{i}{16\pi^2}\frac{x(1-x)}{(M_H^2-m_b^2)} \, (f^W(Q_1) - f^W(Q_0)) \; , 
\label{loopint01}
 \end{eqnarray}
where 
\begin{eqnarray}
f^W(Q) \, \equiv \, \frac{1}{(m_t^2-M_W^2)} \left(
- ln \left(\frac{Q-M_W^2}{M_W^2}\right) 
+ \frac{m_t^2}{(m_t^2-M_W^2)} \left(
dilog(\frac{Q}{m_t^2})  \, -
\,   dilog(\frac{Q}{M_W^2}) \right) \right)
\; .
\label{loopint02}
 \end{eqnarray}
Here, the dilogarithmic function  is in our case  defined as
\begin{equation}
dilog(z) \, = \, \int_1^z dt \frac{ln(t)}{(1-t)} \; = \, Li_2(1-z) \, .
\label{dilog}
\end{equation}
Further, 
 \begin{eqnarray}
Q_1 \, =   \, \frac{1}{x(1-x)}
(m_b^2 \, + \, x(m_t^2 -m_b^2) \, +  y (M_h^2 - m_b^2) + (1-x-y) (M_H^2 -m_b^2))
\; , \nonumber \\
\mbox{and} \; \,  Q_0 \, =  \, 
\frac{1}{x(1-x)}(m_b^2 \, + \, x (m_t^2 - m_b^2) \, +  \, y (M_h^2 - m_b^2))
\, = \, \frac{B}{x(1-x)} \; ,
\label{Qval01}
\end{eqnarray}
where $B$ is defined in (\ref{Rvalue}.)

Now the quantity $S$ in (\ref{loopintJ})  may be split up as : 
\begin{eqnarray}
S^W \, = \, (\frac{i}{16 \pi^2})^2 \, 2 ! \, \int_0^1 \,dx \, \int_0^{(1-x)} \, dy
 \left[f^W(Q_1) - f^W(Q_0) \right] \; = \, S_1^W \, - \, S_0^W \; .
\label{loopint03}
 \end{eqnarray}
Here the quantity $S_1^W$ contains a term $ln(M_H^2)$ corresponding
 to the divergent
term $ln(\Lambda^2)$ in  \cite{Eeg:2016fsy} and (\ref{loopint-h}).
In order to find $S_1^W$ explicitly I use
 the assymptotic property for $Z \rightarrow \infty$
\begin{equation}
dilog(Z) \rightarrow \, - \, \frac{1}{2}\left(ln(Z)\right)^2 \; .
\label{ass}
\end{equation}
Therefore, one obtains for $M_H^2 \gg M_h^2$ 
 \begin{equation}
\left( \, - \, dilog(\frac{Q_1}{M_W^2}) \,  +  \,
 dilog(\frac{Q_1}{m_t^2}) \, \right) \; \rightarrow \;
ln(\frac{m_t^2}{M_W^2}) \cdot \left( ln(\frac{M_H^2}{m_t M_W }) \, + 
\, ln(\sigma) \right) \; \; ,
\label{dilogsAss}
\end{equation}
where
\begin{equation}
\sigma = \frac{(1-x-y)}{x(1-x)} \; .
\label{sigma}
\end{equation}
Then one obtains
\begin{eqnarray}
S_1^W  \, =   (\frac{1}{16\pi^2})^2 \frac{1}{M_H^2 (m_t^2-M_W^2)} \left[
- \left(ln(\frac{M_H^2}{M_W^2}) + \frac{1}{2} \right ) \right. \nonumber \\
+ \left. \frac{m_t^2 ln(m_t^2/M_W^2)}{(m_t^2-M_W^2)} \left(ln(\frac{M_H^2}{m_tM_W})
+ \frac{1}{2} \right) \right] \; ,
\label{loopintFunc}
 \end{eqnarray}
which may be maniplated into 
\begin{eqnarray}
S_1^W  \, =   (\frac{1}{16\pi^2})^2 \frac{(\widetilde{C_{H}})^W \,
\cdot \,  p_W(u_t)}{M_H^2 \, M_W^2}
\; ,
\label{S1}
\end{eqnarray}
where $(\widetilde{C_{H}})^W$ is given 
in (\ref{CH}) and (\ref{CutoffConnection}).

The term $S_0^W$  in (\ref{loopint03}) contains the $ln(R)$ term in 
(\ref{Logsplit}), and is given by
\begin{eqnarray}
S_0^W \, = \, (\frac{i}{16 \pi^2})^2 \, 2 ! \, \int_0^1 \, dx \int_0^{(1-x)} \, dy
\,  f^W(Q_0)  \: .
\label{loopint04}
 \end{eqnarray}
 To see this clear, 
 instead of using (\ref{Logsplit}),  one may use a trick by rewriting
$I_2(R)$ in (\ref{Divloop1}), (\ref{Divloop2}) and (\ref{DivSubloop3}) as 
\begin{equation}
I_2(R) \, = \, 2 \int_R^{\Lambda^2} d\rho \int \frac{\dbar p}{(p^2-\rho)^3}
\; \, .
\label{trick}
\end{equation}
Also, one  observes that 
$R=-x(1-x)(r^2 - Q_0)$.

{\acknowledgments}

 I thank Svjetlana Fajfer for useful comments.

\bibliographystyle{unsrt}

\end{document}